\documentclass{aa}
\usepackage{graphicx}

\begin{document}

\title{A natural explanation for periodic X-ray outbursts
  in Be/X-ray binaries}

\author{A.T.~Okazaki\inst{1,2} \and I.~Negueruela\inst{3}}

\institute{
Faculty of Engineering, Hokkai-Gakuen University, Toyohira-ku,
Sapporo 062-8605, Japan
\and
Institute of Astronomy, Madingley Road, Cambridge CB3 0HA, UK
\and
Observatoire de Strasbourg, 11 rue de l'Universit\'{e}, Strasbourg
F67000 France}

\offprints{okazaki@elsa.hokkai-s-u.ac.jp}

\date{Received ; accepted}

\titlerunning{X-ray outbursts from Be/X-ray binaries}
\authorrunning{Okazaki \& Negueruela}

\abstract{
When applied to Be/X-ray binaries, the viscous decretion disc model,
which can successfully account for most properties of Be stars,
naturally predicts the truncation of the circumstellar disc. The
distance at which the circumstellar disc is truncated depends mainly
on the orbital parameters and the viscosity. In systems with low 
eccentricity, the disc is expected to be truncated at the 3:1
resonance radius, for which the gap between the disc outer radius and
the critical lobe radius of the Be star is so wide that,
under normal conditions, the neutron star cannot accrete enough gas
at periastron passage to show periodic X-ray outbursts
(Type~I outbursts). These systems will display only occasional
giant X-ray outbursts (Type~II outbursts). On the other hand, in
systems with high orbital eccentricity, the disc truncation occurs at
a much higher resonance radius, which is very close to or slightly
beyond the critical lobe radius at periastron unless the viscosity is
very low. In these systems, disc truncation cannot be efficient,
allowing the neutron star to capture gas from the disc at every
periastron passage and display Type~I outbursts regularly. In contrast
to the rather robust results for systems with low eccentricity and
high eccentricity, the result for systems with moderate eccentricity
depends on rather subtle details. Systems in which the disc is
truncated in the vicinity of the critical lobe will regularly display
Type~I outbursts, whereas those with the disc significantly smaller
than the critical lobe will show only Type~II outbursts
under normal conditions and temporary Type~I outbursts when
the disc is strongly disturbed. In Be/X-ray binaries, material will be
accreted via the first Lagrangian point with low velocities relative
to the neutron star and carrying high angular momentum. This may
result in the temporary formation of accretion discs during Type~I
outbursts, something that seems to be confirmed by observations.
}

\maketitle

\section{Introduction}
\label{sec:intro}

Be/X-ray binaries are X-ray sources composed of a Be star and a
neutron star. The high-energy radiation is believed to arise owing to
accretion of material associated with the Be star by the compact
object (see Negueruela \cite{neg98}; see also Bildsten et
al. \cite{bil97}).

A ``Be star'' is an early-type non-supergiant star, which 
at some time has shown emission in the Balmer series 
lines (Slettebak \cite{sle88}, for a review). Both the emission
lines and the characteristic strong infrared excess when compared to
normal stars of the same spectral types are attributed to the presence
of circumstellar material in a disc-like geometry. The causes that
give rise to the disc are not well understood. Different mechanisms
(fast rotation, non-radial pulsation, magnetic loops) have been
proposed, but it is still unclear whether any of them can explain the
observed phenomenology on its own. The discs are rotationally
dominated and motion seems to be quasi-Keplerian (Hanuschik
\cite{han96}). However, some kind of global outflow is needed to
explain the X-ray emission from Be/X-ray binaries (Waters et
al. \cite{wat88}).

Some Be/X-ray binaries are persistent X-ray sources (see Reig \& Roche
\cite{rr99}), displaying low luminosity ($L_{{\rm x}} \sim 
10^{34}\:{\rm erg}\,{\rm s}^{-1}$) at a relatively constant level
(varying by up to a factor of $\sim 10$). On the other hand, most
known Be/X-ray binaries (though this is probably a selection effect)
undergo periods in which the X-ray luminosity suddenly increases by a
factor $\ga 10$ and are termed Be/X-ray transients.

Be/X-ray transients fall within a relatively narrow area in the 
$P_{{\rm orb}}$/$P_{{\rm spin}}$ diagram (see Corbet \cite{cor86};
Waters \& van Kerkwijk \cite{wk89}), indicating that some mechanism
must be responsible for the correlation. Those systems with
fast-spinning neutron stars do not show pulsed X-ray emission during
quiescence (though non-pulsed radiation  could be caused by accretion
on to the magnetosphere) because of the centrifugal inhibition of
accretion (Stella et al. \cite{swr86}). Systems with more slowly
rotating pulsars show X-ray emission at a level $L_{{\rm x}} \la
10^{35}$ erg s$^{-1}$ when in quiescence. Transients show two
different kinds of outbursts:

\begin{itemize}
\item X-ray outbursts of moderate intensity ($L_{{\rm x}} 
   \approx 10^{36} - 10^{37}$ erg s$^{-1}$) occurring in series
   separated by the orbital period (Type I or normal), generally (but
   not always) close to the time of periastron passage of the neutron
   star. In most cases, the duration of these outbursts seems to be
   related to the orbital period.
\item Giant (or Type II) X-ray outbursts ($L_{{\rm x}} 
   \ga 10^{37}$ erg s$^{-1}$) lasting for several weeks or even
   months. Generally Type II outbursts start shortly after periastron
   passage, but do not show any other correlation with orbital
   parameters (Finger \& Prince \cite{fp97}). In systems like
   \object{4U\,0115+63} the duration of the Type II outbursts seems to
   be to some degree correlated with its peak intensity, but in
   \object{A\,0535+262} Type I outbursts may be as long as much
   brighter Type II outbursts (Finger et al. \cite{fin96a}).
\end{itemize}

\section{Radial outflows vs. quasi-Keplerian discs}
\label{sec:windmodel}

Attempts at modelling the X-ray luminosities of Be/X-ray transients
during outbursts have made use of a simple wind accretion model, in
which the neutron star accretes from a relatively fast radial
outflow. The disc of the central Be star is supposed to have a
power-law density distribution
\begin{equation}
\rho(r) = \rho_{0} \left( \frac{r}{R_{*}} \right)^{-n}
\end{equation}
where $\rho_{0}$ is the density at the stellar surface and $R_{*}$ is
the radius of the star. This results in a power velocity law of the
form
\begin{equation}
v(r) = v_{0} \left( \frac{r}{R_{*}} \right)^{n-2}
\end{equation}
where the values of $v_{0}$ and $n$ have to be determined
observationally (see Waters et al. \cite{wat89} and references
therein). The rotational velocity of the outflow takes the form
\begin{equation}
v_{{\rm rot}}(r) = v_{{\rm rot,0}} \left( \frac{r}{R_{*}}
\right)^{-\alpha}
\end{equation}
with $0.5 \le \alpha \le 1$ (respectively the Keplerian case and
conservation of angular momentum). 

Accretion is considered to follow the classical Bondi-Hoyle-Littleton 
(BHL) approximation. The most important parameter, the relative
velocity between the outflow and the neutron star, can be written as
\begin{equation}
v^{2}_{{\rm rel}} = \left( v - V_{{\rm rad}}\right)^{2} + 
 \left( v_{{\rm rot}} - V_{{\rm rot}}\right)^{2}
\end{equation}
where $V_{{\rm rad}}$ and $V_{{\rm rot}}$ are the radial and
tangential components of the orbital velocity of the neutron star. 
Material is supposed to be accreted when it is within a capture radius
defined as
\begin{equation}
r_{{\rm c}} = 2GM_{{\rm x}} v^{-2}_{{\rm rel}}
\end{equation}
where $M_{{\rm x}}$ is the mass of the neutron star and $G$ is the
gravitational constant. The X-ray luminosity in the BHL approximation
can then be expressed as
\begin{equation}
L_{{\rm x}} = 4\pi G^{3}M^{3}_{{\rm x}}R^{-1}_{{\rm x}}
v^{-4}_{{\rm rel}}F_{m} \propto \rho v^{-3}_{{\rm rel}}
\end{equation}
where $R_{{\rm x}}$ is the radius of the neutron star and $F_{m}=\rho
v_{{\rm rel}}$ is the mass flow. In order to explain the wide range of
observed X-ray luminosities, large changes in the value of the radial
velocity have to be invoked. For example, Waters et al. (\cite{wat89})
deduced that the relative velocity was $v_{{\rm rel}} \approx
300\:{\rm km}\,{\rm s}^{-1}$ during a Type I outburst of
\object{V\,0332+53} in 1983, while it was $\ll 100\:{\rm km}\,
{\rm s}^{-1}$ during a Type II outburst in 1973.

The use of the BHL approximation implies a number of simplifying
assumptions which are not always easy to justify. For example, it
neglects any effect of the mass-losing star, which for periastron
distances of $\sim 10\:R_{*}$ and mass ratios $q \sim 0.1$ seems to be
an excessive simplification. In addition, while the use of the
accretion radius formalism is adequate for an accreting object
immersed in a medium, its application to an outflow characterised by a
relatively small scale-height ($H$), such as a Be disc, is dubious.

Moreover, when the low outflow velocities required to explain Type II
outbursts are considered, the formalism breaks down completely, since
the capture radius becomes far too large to have any physical
meaning. For example, for $v_{{\rm rel}} \simeq 20\:{\rm km}\,{\rm
  s}^{-1}$, $r_{{\rm c}} \simeq 9.6\times10^{11}\:{\rm m} \approx
1400\:R_{\sun}$, which is one order of magnitude larger than the 
binary separation. A crude way round this problem is to consider that
the radius of the effective Roche lobe of the neutron star 
$r_{{\rm R}}$ should be used instead of $r_{{\rm c}}$ whenever the
calculated value is larger than  $r_{{\rm R}}$ (e.g., Ikhnasov
\cite{ikh01}). In spite of these shortcomings, the model has been
repeatedly used in an attempt to model lightcurves of Be/X-ray
binaries (Raguzova \& Lipunov \cite{rl98}; Reig et al. \cite{rei98})
with only moderate success.

Beyond the purely formal aspects, one obvious difficulty for the model 
is the fact that many Be/X-ray binaries (e.g., \object{A\,0535+262})
show low-luminosity X-ray emission when they are not in outburst. It
is believed that all Be/X-ray binaries for which centrifugal
inhibition of accretion is not effective display this emission. Motch
et al. (\cite{mot91}) detected \object{A\,0535+262} on several
occasions at luminosities of $\approx 2\times10^{35}\:{\rm erg}\,{\rm
  s}^{-1}$. In order to explain this luminosity within the framework
described above and taking into account that optical and infrared
observations do not show any sign of the large variations that would
be associated with a change of several orders of magnitude in the
density of material, enormous relative velocities (of the order of
$\sim 10^{4}\:{\rm km}\,{\rm s}^{-1}$) are needed.

One further complication comes from the fact that Be/X-ray binaries
spend most of their time in the quiescent state described in the
previous paragraph and only occasionally show series of outbursts. The
model does not offer any explanation as to why there could be a change
from quiescence to outburst, unless again very large and sudden
changes in the density and velocity of the flow are assumed. Given the
changes in relative velocities needed to account for the observed
range of X-ray luminosities and the lack of any physical mechanism
that could explain them, it is clear that direct accretion from a
wind-like outflow is not the best approximation to the way in which
matter is fed on to the neutron star.

But the major objection to the model is simply the fact that there 
is no observational evidence whatsoever supporting the 
existence of such fast outflows. All observations of Be stars imply 
bulk outflow velocities smaller than a few ${\rm km}\,{\rm s}^{-1}$
(Hanuschik \cite{han00}).
The evidence for rotationally dominated quasi-Keplerian discs around Be 
stars is overwhelming (see Hanuschik et al. \cite{hanal96}; Hummel \&
Hanuschik \cite{hum97}; Okazaki \cite{oka97}), especially owing to the
success of the one-armed global oscillation model to explain V/R
variability in the emission lines of Be stars (Kato
 \cite{kat83}; Okazaki \cite{oka91}, \cite{oka96}; Papaloizou et
al. \cite{papa92}; Hummel \& Hanuschik \cite{hum97}).

Therefore it seems necessary to attempt an explanation of the outburst
behaviour of Be/X-ray binaries that does not imply large outflow
velocities.

\section{The viscous disc model}
\label{sec:viscmodel}

Whatever the mechanism originating the Be phenomenon, the model which
at present appears more applicable to explaining the discs surrounding
Be stars is the viscous decretion disc model (Lee et al. \cite{lee91};
see Porter \cite{por99} and Okazaki \cite{oka01} for detailed
discussion). In this scenario, angular momentum is transferred from
the central star by some mechanism still to be determined 
(perhaps associated with non-radial pulsations) to the inner edge of
the disc, increasing its angular velocity to Keplerian. Viscosity
then, operating in a way opposite to an accretion disc, conducts
material outwards. In this scenario, material in the disc moves in
quasi-Keplerian orbits and the radial velocity component is highly
subsonic until the material reaches a distance much larger than the
line-emitting regions (Okazaki \cite{oka01}). The outflow is very
subsonic for the distances at which neutron stars orbit in close
Be/X-ray transients $\left(v(r) <  1{\rm km}\,{\rm s}^{-1}\right)$ and
still subsonic for the orbital sizes of all Be/X-ray binaries for
which there is an orbital solution. The viscous decretion disc model
successfully accounts for most of the observational characteristics of
Be discs.

Negueruela \& Okazaki (\cite{no01}, henceforth Paper~I) have modelled
the disc surrounding the Be primary in the Be/X-ray transient
\object{4U\,0115+63} as a viscous decretion disc and found that the
tidal interaction of the neutron star naturally produces the
truncation of the circumstellar disc, as it does for accretion
discs in close binaries (Paczy\'nski \cite{pac77}).
The result of Negueruela \& Okazaki (\cite{no01}) is in agreement with
the results of Reig et al. (\cite{rei97}), who showed that there is a
correlation between the orbital size and the maximum equivalent width
of H$\alpha$ ever observed in a system. Even though it is clear that
the equivalent width of H$\alpha$ is not an effective measurement of
the size of the disc owing to several effects (see, for example,
Negueruela et al. \cite{negal98}), the maximum equivalent width ever
observed becomes a significant indicator if the system has been
monitored during a period which is long in comparison with the typical
time-scale for changes in the disc (which, if viscosity is dominant,
should be only a few months). Therefore the result of Reig et
al. (\cite{rei97}) clearly indicates that the neutron star has some
sort of effect on the size of the disc.

In this paper we apply the model presented in Paper I 
to several Be/X-ray transients for which orbital solutions exist and 
investigate how the truncation radius depends on different orbital 
parameters. 

\section{Model description and limitations}
\label{sec:trunc}

The model developed in Paper I describes a binary system in which a
primary Be star of mass $M_{*}$ and radius $R_{*}$ is orbited by a
neutron star of mass $M_{{\rm x}} = 1.4\:M_{\sun}$ which moves in an
orbit of eccentricity $e$ and period $P_{\rm orb}$. The Be star is
assumed to be surrounded by a near-Keplerian disc which is primarily
governed by pressure and viscosity. For simplicity, the disc is
assumed to be isothermal and Shakura-Sunyaev's viscosity prescription
is adopted.

In such a disc, angular momentum is added to the disc by the viscous
torque, whereas it is removed from the disc by the resonant torque
exerted by the neutron star companion, which becomes non-zero only at
radii where the ratio between the angular frequency of disc rotation
and the angular frequency of the mean binary motion is a rational
number. As a result, the disc decretes outward owing to the  transfer
of angular momentum by viscosity until the resonant torque becomes
larger than the viscous torque at a resonant radius. 

Therefore, the criterion for the disc truncation at a given resonance
radius is written as
\begin{equation}
   T_{\rm vis} + T_{\rm res} \le 0,
   \label{eqn:criterion}
\end{equation}
where $T_{\rm vis}$ and $T_{\rm res}$ are the viscous torque and the
resonant torque, respectively.

The viscous torque $T_{\rm vis}$ is written as
\begin{equation}
   T_{\rm vis} = 3 \pi \alpha GM_{*} \sigma r \left( {H \over r}
                 \right)^2
   \label{eqn:t_vis}
\end{equation}
(Lin \& Papaloizou \cite{lp86}), where $\sigma$ is the surface density
of the disc, $\alpha$ is the Shakura-Sunyaev viscosity parameter and
$H$ is the vertical scale-height of the disc given by
\begin{equation}
   {H \over r} = {c_{\rm s} \over {V_{\rm K}(R_*)}}
                 \left({r \over R_*} \right)^{1/2}
   \label{eqn:h}
\end{equation}
for the isothermal disc.
Here, $c_{\rm s}$ is the isothermal sound speed and $V_{\rm K}(R_*)$
is the Keplerian velocity at the stellar surface.
In the systems we will discuss later, $c_{\rm s}/V_{\rm K}(R_*)$
ranges $3.1-4.1 \cdot 10^{-2}(T_{\rm d}/T_{\rm eff})^{1/2}$, where
$T_{\rm d}$ and $T_{\rm eff}$ are the disc temperature and the effective
temperature of the Be star, respectively. Note that the viscous torque
is proportional to the disc temperature.

The resonant torque $T_{\rm res}$ is calculated by using Goldreich \& 
Tremaine's (\cite{gt79}, \cite{gt80}) torque formula, after
decomposing the binary potential $\Phi$ into a double Fourier series
as
\begin{eqnarray}
   \Phi (r, \theta, z) &=& -{{GM_{*}} \over r}
   -{{GM_{\rm x}} \over{[r^2+r_2^2-2rr_2 \cos(\theta-f)]^{1/2}}}
   \nonumber \\
   && +{{GM_{\rm x}r} \over r_2^2} \cos(\theta-f) \nonumber \\
   &=& \sum_{m,l} \phi_{ml} \exp[i(m \theta - l \Omega_B t)],
   \label{eqn:pot1}
\end{eqnarray}
where $r_2$ is the distance of the neutron star from the primary, $f$
is the true anomaly of the neutron star, $m$ and $l$ are the azimuthal
and time-harmonic numbers, respectively, and $\Omega_B =
[G(M_{*}+M_{\rm x})/a^3]^{1/2}$ is the mean motion of the binary with
semimajor axis $a$. The pattern speed of each potential component is
given by $\Omega_p=(l/m)\Omega_B$. The third term in the right hand
side of the first equation is the indirect potential arising because
the coordinate origin is at the primary.

For each potential component, there can be three kinds of resonances,
i.e., the outer and inner Lindblad resonances at radii $[ (m \pm
1)/l]^{2/3}(1+q)^{-1/3} a$, where $\Omega_p=\Omega \pm \kappa/m$, and
a corotation resonance (CR) at the radius $(m/l)^{2/3}(1+q)^{-1/3} a$,
where $\Omega_p=\Omega$. Here, $\kappa$ is the epicyclic frequency and
the upper and lower signs correspond to the outer Lindblad resonance
(OLR) and inner Lindblad resonance (ILR), respectively.
In circumstellar discs, however, the resonant torque from the inner 
Lindblad resonance, which is given by
\begin{equation}
   (T_{ml})_\mathrm{ILR} = -{{m(m-1)\pi^2 \sigma (\lambda + 2m)^2
            \phi_{ml}^2}
            \over {3l^2 \Omega_B^2}},
   \label{eqn:tml_ilr}
\end{equation}
where $\lambda = d \ln \phi_{ml}/d \ln r$, 
always dominates the resonant torques from the corotation resonance
and the outer Lindblad resonance. The resonant torque at a given
resonance radius is then given by
\begin{eqnarray}
   T_{\rm res} &=& \sum_{ml}(T_{ml})_{\rm ILR}
   +\sum_{m^{'}l^{'}}(T_{m^{'}l^{'}})_{\rm
   OLR}+\sum_{m^{''}l^{''}}(T_{m^{''}l^{''}})_{\rm CR}
   \nonumber \\
   &\simeq& \sum_{ml}(T_{ml})_{\rm ILR}.
   \label{eqn:t_res}
\end{eqnarray}
Since high-order potential components contribute little to the total
torque, the summation in Eq.~(\ref{eqn:t_res}) is safely taken over
several lowest-order potential components which give the same radius.

For a given set of stellar and orbital parameters and the disc
temperature, criterion~(\ref{eqn:criterion}) at a given resonance is
met for $\alpha$ smaller than a critical value $\alpha_{\rm crit}$ and
we assume that the disc is truncated at the resonance if $\alpha <
\alpha_{\rm crit}$.

In Paper I it was shown that for any subsonic outflow, the drift
time-scale $\tau_{\rm drift} \sim \Delta r/v_r \sim
{\cal M}_r^{-1}(\Delta r/H) \Omega^{-1}$ was considerably
longer than a typical truncation time-scale $\tau_{\rm trunc} \sim
(\alpha/\alpha_{\rm crit})\tau_{\rm vis} \sim \alpha_{\rm
  crit}^{-1}(\Delta r/H)^2\Omega^{-1}$, where $\Delta r$ is the gap
size between the truncation radius and the radius where the gravity by
the neutron star begins to dominate, and $v_r$ and ${\cal M}_r$
are the radial velocity and Mach number, respectively. As a
consequence, the tidal and resonant interaction with the neutron star
led to disc truncation.

One important consequence of the above is that the discs surrounding
the primaries in Be/X-ray binaries cannot reach a steady state. Most
of the outflowing material loses angular momentum and falls back
towards the central star. As a consequence of the interaction with the
material which is coming outwards from the inner regions, it is likely
that the disc becomes denser and the density distribution in the
radial direction becomes flatter with increasing time. We note that we
do not expect the truncation effect to be one hundred per cent
efficient. This is not only owing to theoretical considerations (see
below), but probably required by the existence of pulsed
low-luminosity X-ray emission during quiescence.

In the formulation above only torques integrated over the whole orbit
are considered. Given the large eccentricities observed in Be/X-ray
binaries (mostly larger than 0.3 and sometimes approaching 0.9), the 
gravitational effect of the neutron star is very strongly 
dependent on the orbital phase. This means that in such systems the
disc radius and the truncation radius are also phase-dependent. The
disc would shrink at periastron, at which the truncation radius 
becomes smallest, while it would spread when the neutron star is far
away and its gravitational effect is not felt so strongly. The spread
of the disc would continue until the truncation radius becomes
smaller than the disc radius at some phase before the next periastron
passage. This variation in disc radius would be larger for a larger
eccentricity. For a longer orbital period, the effect would be yet
stronger. Given that, for some systems considered (see Section
\ref{sec:sources}), the model truncation radius is close to the
critical lobe radius at periastron, this will provide a mechanism by
which disc material can reach the neutron star (see also
Fig.~\ref{fig:scenarios} for scenarios for two families of Type~I
outbursts).

\section{Modelling the Be/X-ray binaries}
\label{sec:sources}

Table~\ref{tab:galax} shows a list of known X-ray binaries which have 
exhibited outbursting behaviour or some sort of orbital modulation in
their X-ray lightcurve.
The top panel contains Be/X-ray binaries with an identified Be
counterpart which have displayed Type~I X-ray outbursts, i.e., a
series of outbursts separated by their orbital period (note that in
many cases there is no orbital solution for the system and the
recurrence period of the outbursts is taken to be the orbital
period). The middle
panel shows  other systems without identified optical counterparts
whose X-ray behaviour marks them out as Be/X-ray binaries. Finally the
bottom panel contains a few other Be/X-ray binaries with identified
counterparts whose X-ray behaviour deviates slightly from what is
considered typical (and which will be discussed individually in
Section~\ref{sec:discu}).

In this section, we apply our model to systems for which exact orbital
solutions have been deduced from the analysis of Doppler shifts in the
arrival times of X-ray photons. They include five Be/X-ray binaries in
the top panel (\object{V\,0332+53}, \object{A\,0535+262},
\object{GRO\,J1008$-$57}, \object{2S\,1417$-$624}, and
\object{EXO\,2030+375}) and a likely Be/X-ray binary in the middle
panel (\object{2S\,1845$-$024}). These systems are discussed
individually in the following subsections. The case of the Be/X-ray
transient \object{4U\,0115+63} has been carefully discussed in Paper I
and therefore it will not be included here.  

\begin{table*}[ht]
\caption{List of known X-ray binaries to which our model could be
  applied. The top panel contains Be/X-ray binaries with an identified
  counterpart which have displayed Type I X-ray outbursts. The middle
  panel shows  other systems without identified optical counterparts
  likely to be Be/X-ray binaries. The bottom panel contains
  some other Be/X-ray
  binaries with identified counterparts in which the nature of the
  outbursts is not clear yet (all these systems are discussed in the
  text). Basic orbital parameters are listed.\newline
  \hspace*{1em}
 Orbital periods marked with `$^{*}$' represent the recurrence time of
  X-ray outbursts; other orbital periods are derived from exact
  orbital solutions. Spectral types have only been included for those
  systems in which they are derived from high signal-to-noise ratio
  spectra in the classification region. Objects in which the
  eccentricity is marked as `large' have no orbital solutions, but
  available data imply $e>0.5$. References are only given if 
  data were not included in Negueruela (\cite{neg98}).}
\begin{center}
\begin{tabular}{lccccccc}
\hline
Name & Optical &  Spectral &$P_{{\rm s}}$(s) & $P_{{\rm orb}}$(d) & $e$\\
& Counterpart & Type & & & \\
\hline
\object{4U\,0115+63}4 & \object{V635 Cas} & B0.2V$^{\mathrm{a}}$ & 3.6 & 24.3 & 0.34\\
\object{V\,0332+53} & \object{BQ Cam} & O8.5V$^{\mathrm{b}}$ & 4.4 & 34.2 & 0.31 \\
\object{A\,0535+262} & \object{V725 Tau} & B0III& 103 & 111 & 0.47\\
\object{RX J0812.4$-$3114} & \object{LS 992} & B0.2III$^{\mathrm{c}}$ & 31.9 & 81$^{*}$$^{\mathrm{d}}$& $-$\\
\object{GS\,0834$-$430} & star$^{\mathrm{e}}$ & $-$ & 12.3 & 105.8 & $0.1<e<0.17$$^{\mathrm{f}}$\\
\object{GRO\,J1008$-$57} & star& $-$ &93.5&247.5$^{*}$& 0.66$^{*}$$^{\mathrm{g}}$ \\
\object{4U\,1145$-$619} & \object{V801 Cen} & B0.2III$^{\mathrm{h}}$ & 292 & 187$^{*}$ &
large\\
\object{4U\,1258$-$61}  & \object{V850 Cen} & B0.7V$^{\mathrm{h}}$ & 272 &132.5$^{*}$& large\\
\object{2S\,1417$-$624} &  star & $-$ & 17.6 & 42.1& 0.45\\
\object{XTE\,J1946+274} & star & $-$ & 15.8 & 172$^{\mathrm{i}}$& $-$\\
\object{EXO\,2030+375} & star & $-$ & 41.7 & 46.03  & 0.41$^{\mathrm{j}}$ \\
\hline
\object{XTE\,J1543$-$568} & $-$ & $-$ & 27.1 & 76.6 & $<0.03$$^{\mathrm{k}}$ \\
\object{2S\,1845$-$024} & $-$ & $-$ & 94.3 & 242.2 & 0.88$^{\mathrm{l}}$ \\
\object{GRO\,J2058+42} & $-$ & $-$ & 198 & 110$^{*}$$^{\mathrm{m}}$ & $-$ \\
\hline
\object{A\,0535$-$668} & star & B0.5III$^{\mathrm{n}}$ & 0.07 & 16.7$^{*}$ & large\\ 
\object{A\,0726$-$26} & \object{V441 Pup} & O8.5V& 103.2 & 35$^{*}$  & $-$\\
\object{A\,1118$-$616} & \object{Wray 977} &O9.5V& 406.5 & $-$ & $-$\\
\object{Cep X$-$4} & star & B1V?& 66.3 & $-$ & $-$ \\
\object{SAX\,J2239.3+6116} & star & $-$ & $-$ & 262$^{*}$$^{\mathrm{o}}$ & $-$\\
\hline
\end{tabular}
\begin{tabbing}
circusvariousautomati\=$^{\mathrm{a}}$ Negueruelas \& Okazakir
(2001)\=$^{\mathrm{a}}$ Negueruelam et alt., in prep.\=Rumpelstinski and
friends (1233)\=\kill
\>$^{\mathrm{a}}$ Negueruela \& Okazaki (\cite{no01})\> $^{\mathrm{b}}$
Negueruela et al. (\cite{negal99})\>$^{\mathrm{c}}$ Reig et al. (\cite{rei01})\\
\>$^{\mathrm{d}}$ Corbet \& Peele (\cite{cp00}) \>$^{\mathrm{e}}$ Israel et
al. (\cite{isr00})\>  $^{\mathrm{f}}$ Wilson et al. (\cite{wil97})\\
\>$^{\mathrm{g}}$ M. Scott, priv. comm.\> $^{\mathrm{h}}$ Negueruela, in
prep.\>$^{\mathrm{i}}$ Wilson et al., in prep.\\
\>$^{\mathrm{j}}$ Wilson et al. (\cite{wil01})\> $^{\mathrm{k}}$ in't Zand et al. (\cite{zan01})\> $^{\mathrm{l}}$ Finger et al. (\cite{fin99})\\ 
\>$^{\mathrm{m}}$ Wilson et al. (\cite{wil98})\> $^{\mathrm{n}}$ Negueruela \& Coe
(\cite{nc01}) \> $^{\mathrm{o }}$in't Zand et
al. (\cite{zan00})\>\\ 
\end{tabbing}
\end{center}
\label{tab:galax}
\end{table*}

As mentioned in the previous section, we adopt Shakura-Sunyaev's
viscosity prescription, in which the viscosity parameter $\alpha$ is a
free parameter, and assume the Be disc to be isothermal. 
In what follows, we adopt $T_{\rm d}=\frac{1}{2}T_{\rm eff}$.
Note that our assumption of the disc temperature is consistent 
with the results by Millar and Marlborough (\cite{mm98}, \cite{mm99}),
who computed the distribution of the disc temperature within
$100\,R_{*}$ around the B0 star $\gamma$~Cas and the B8--9 star 1~Del
by balancing at each position the rates of energy gain and energy loss
and found that the disc is roughly isothermal at a temperature about
half the effective temperature of the star.

When adopting a particular model for
a system, the main source of uncertainty comes from the choice of mass
for the primary, a parameter which can only be guessed from the
spectral type. We find two main difficulties. In some cases, the
spectral type of the primary is not well determined. Moreover, there
is some evidence that fast rotators may be moderately over-luminous
for their masses (see Gies et al. \cite{gie98}). For this reason, in
general we have taken masses slightly lower than those given in the
calibration of Vacca et al. (\cite{vac96}). The spectral distribution
of primaries of Be/X-ray binaries (Negueruela \cite{neg98}) is
strongly peaked at B0. Therefore for systems without exact
determination of the spectral type, we have calculated models
corresponding to B0V and B0III primaries. In any case, our results
show that the exact mass of the primary is not one of the main factors
in the X-ray behaviour of the sources.

\subsection{\object{V\,0332+53}}
\label{sec:V0332+53}

This transient pulsar has an orbital period $P_{{\rm orb}}=
34.25\,{\rm d}$ with a relatively
low eccentricity of $e =0.31$ (Stella et al. \cite{stel85}).
The optical component of this system is an unevolved star in the
O8\,-\,9 range (Negueruela et al. \cite{negal99}). For O8.5V stars
Vacca et al. (\cite{vac96}) give an spectroscopic mass $M_{*} =
23.6\,M_{\sun}$ and a theoretical mass $M_{*} = 28\,M_{\sun}$. Here we
will assume as a conservative model the  lower limit of $M_{*} =
20\,M_{\sun}$ and $R_{*}=8.8\,R_{\sun}$. The mass function $f(M)
=0.1$ then implies $\sin i = 0.17$, resulting in  an orbital
separation $a \simeq 130\,R_{\sun}$ ($r_{{\rm per}}\approx10\,R_{*}$). 

This system has been rarely detected in X-rays. A Type II outburst
was observed in 1973 (see Negueruela et al. \cite{negal99} for
references). Ten years later, it was observed during a series of three
Type I outbursts. Finally it was observed during another Type II
outburst in 1989. Between 1991 and 2000 it has not been detected by
either the BATSE instrument on board the {\em ComptonGRO} satellite or
the All Sky Monitor on board {\em RossiXTE} and it is believed to be
in a dormant state (Negueruela et al. \cite{negal99}). 

\begin{figure}[t]
 \resizebox{\hsize}{!}{\includegraphics{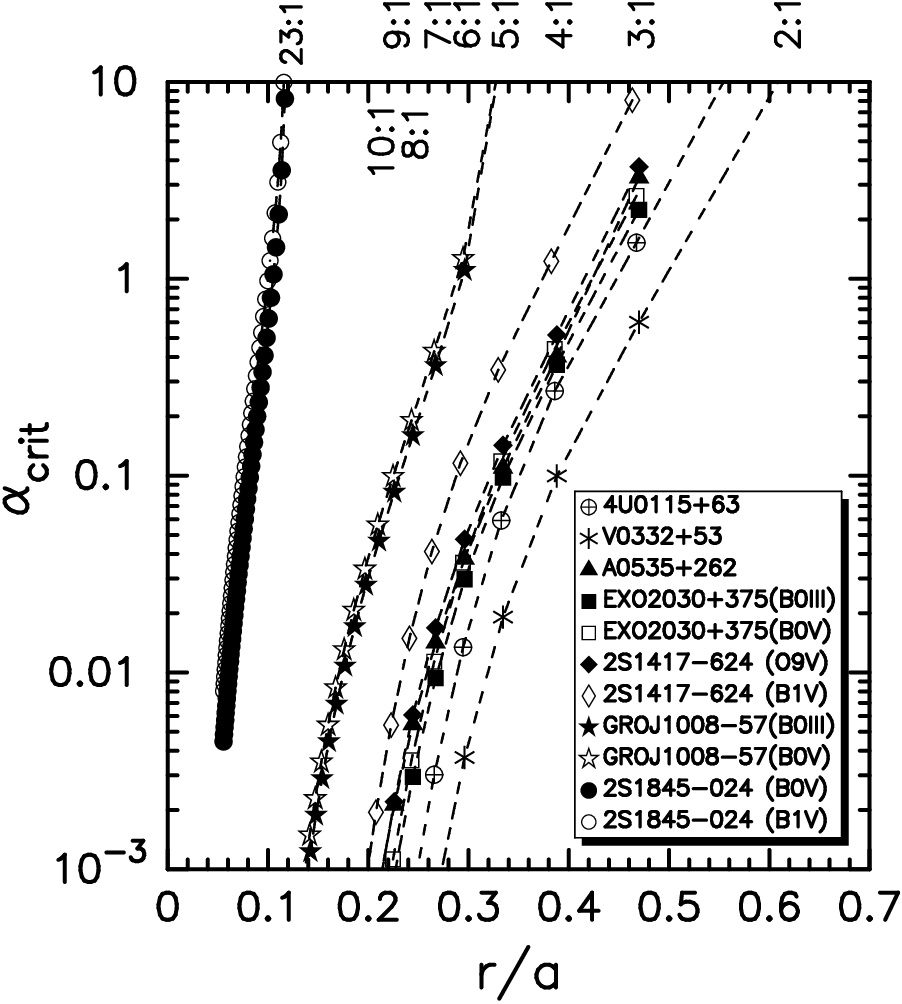}}
   \caption{Critical values of $\alpha$ at some resonance radii for
     systems discussed in the text. Annotated in the figure are the
     locations of the $n:1$ commensurabilities of disc and mean binary
     orbital frequencies. $T_{\rm d} = \frac{1}{2}T_{\rm eff}$ is
     adopted for all models. For other disc temperatures,
     $\alpha_{\rm crit}$ should be multiplied by a factor of $T_{\rm
     eff}/2T_{\rm d}$.}
   \label{fig:alpha_c}
\end{figure}

In Fig.~\ref{fig:alpha_c}, we plot $\alpha_{\rm crit}$ at the $n:1$
resonance radii for those Be/X-ray binaries which will be discussed in
this section. The resonant torques at the $n:1$ radii are stronger
than those at radii with other period commensurabilities located
nearby. Note that we have adopted a particular disc temperature,
$T_{\rm d} = \frac{1}{2}T_{\rm eff}$, for each stellar model. For
other disc temperatures, $\alpha_{\rm crit}$ should be multiplied by a
factor of $T_{\rm eff}/2T_{\rm d}$, taking account of the fact that
the viscous torque is proportional to $T_{\rm d}$.

Fig.~\ref{fig:alpha_c} shows that the Be disc in \object{V\,0332+53} is
expected to be truncated at the 3:1 resonance radius ($r_{\rm t}/a
\simeq 0.47$, where $r_{\rm t}$ is the truncation radius) for $0.099
\la \alpha \la 0.60$ and at the 4:1 resonance radius ($r_{\rm t}/a
\simeq 0.39$) for $0.019 \la \alpha \la 0.099$.

\begin{figure*}[ht]
   \sidecaption
   \includegraphics[width=12cm,clip]{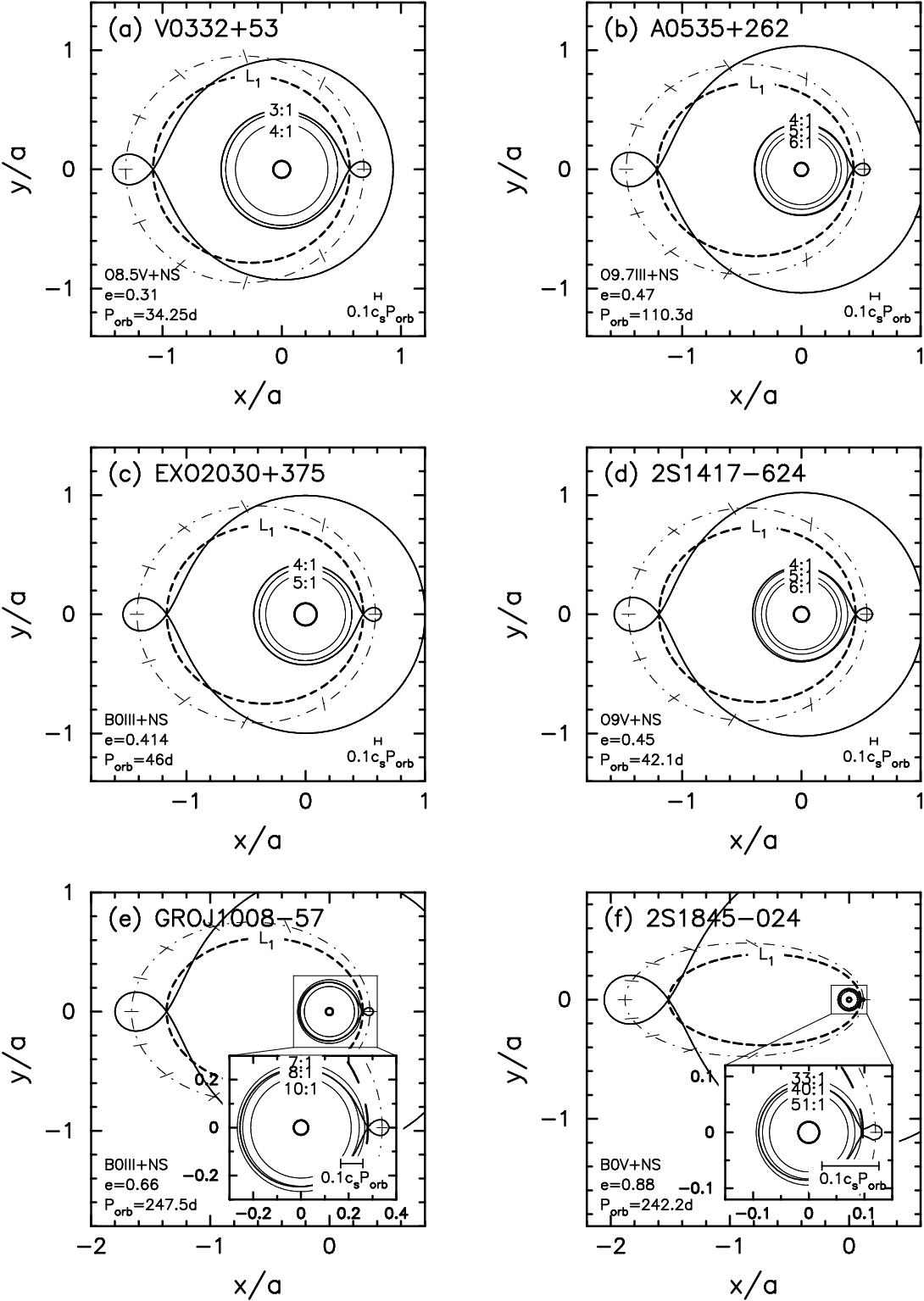}
   \caption{ Description of the orbital model adopted for each system,
   in the reference system centred on the Be star. The dash-dotted
   line represents the orbit of the neutron star. The thick dashed
   line represents the position of the first Lagrangian point ($L_1$)
   around the orbit. The solid thin lines are the critical
   lobes of the two stars at apastron and periastron (the position
   of the neutron star is marked with a cross). The labelled solid
   lines represent the locations of the $n:1$ commensurabilities of
   disc and binary orbital periods at which truncation occurs for
   $\alpha = 0.3$, 0.1, and 0.03 (starting from the outside). The
   distance scale corresponding to $0.1 c_{\rm s} P_{\rm orb}$ is
   shown at the lower-right corner of each panel.}
   \label{fig:roche}
\end{figure*}

It is interesting to see how close the truncation radius is to the
size of the critical lobe at periastron. Fig.~\ref{fig:roche} shows
orbital models for the systems discussed in this section. The
potential $\psi$ describing the effects of the gravitational and
centrifugal forces on the motion of test particles orbiting the Be
star is given by
\begin{equation}
   \psi (r, \theta, z) = \Phi (r, \theta, z)
   -{1 \over 2} \Omega^2(r) r^2,
   \label{eqn:roche}
\end{equation}
where $\Phi$ is the potential defined by Eq.~(\ref{eqn:pot1}). Also
shown is the distance scale corresponding to $0.1 c_{\rm s} P_{\rm
 orb}$. This scale should be taken as an upper limit of the distance
over which the disc outer radius can spread out during one orbital
period, because the outflow velocity in Be discs is certainly much
smaller than a few km\,s$^{-1}$.

From Fig.~\ref{fig:alpha_c} and Fig.~\ref{fig:roche}(a), we note that
the truncation radii of the Be disc in \object{V\,0332+53} for $\alpha
\la 0.60$ is smaller than the critical lobe size at periastron. In
other words, the Be disc never fills the critical lobe even at
periastron unless the viscosity is very high ($\alpha \ga 0.6$).

Table~\ref{tab:gap} gives the gap size 
$\Delta r = r_{\rm crit}-r_{\rm t}$,
where $r_{\rm crit}$ is the mean radius of the critical lobe at
periastron, together with stellar parameters adopted and the resulting
truncation radii. Table~\ref{tab:gap} also gives lower limits for the
drift timescale, $(\tau_\mathrm{drift}/P_\mathrm{orb})_\mathrm{min}$,
which is given by $(\tau_\mathrm{drift}/P_\mathrm{orb})_\mathrm{min} 
\sim \Delta r/[(v_r)_\mathrm{max} P_\mathrm{orb}]
\sim \Delta r/(0.1 c_\mathrm{s} P_\mathrm{orb})$.
For a system with $(\tau_\mathrm{drift}/P_\mathrm{orb})_\mathrm{min} >
1$, the gap is wide and the truncation is efficient, while the spread
of the disc can make the truncation inefficient for a system with
$(\tau_\mathrm{drift}/P_\mathrm{orb})_\mathrm{min} < 1$.

Given that we expect the viscosity in the disc to be of the order of
$\alpha \la 0.1-1$, the Be disc in \object{V\,0332+53} is likely to be
truncated at the 3:1 resonance radius or the 4:1 resonance radius. If 
the disc is truncated at the 3:1 radius, which is close to the mean
critical radius at periastron [$\Delta r/a \simeq 0.03$
and $(\tau_\mathrm{drift}/P_\mathrm{orb})_\mathrm{min} \simeq 0.56$],
a small perturbation easily causes the
outermost part of the disc to fall into the gravitational well of the
neutron star. However, the X-ray history of this system described
above suggests that the system shows no Type~I X-ray outburst in its
normal state. Therefore, we expect that the disc in
\object{V\,0332+53} has a viscosity parameter
(slightly) less than 0.1 and/or a temperature (slightly) lower than
$\frac{1}{2} T_{\rm eff}$. Under those conditions, it will be
truncated at the 4:1 resonance radius
[$\Delta r/a \simeq 0.11$ and 
$(\tau_\mathrm{drift}/P_\mathrm{orb})_\mathrm{min} \simeq 2.1$].

\begin{table*}[ht]
\caption{Stellar parameters adopted for the systems modelled (those
   with known orbital parameters) and resulting critical lobe radius 
   $r_{\rm crit}$, gap size $\Delta r$, and
   lower limit for $\tau_\mathrm{drift}/P_\mathrm{orb}$
   for different values of the viscosity parameter.}
\begin{tabular}{lccrclllllll}
   \hline
   &&&&&&
   \multicolumn{2}{c}{$\alpha=0.03$} &
   \multicolumn{2}{c}{$\alpha=0.1$} &
   \multicolumn{2}{c}{$\alpha=0.3$} \\
   \cline{7-12}
   \multicolumn{1}{c}{Name} &
   \begin{tabular}{@{}c@{}}
       Spectral\\ Type
   \end{tabular} &
   \begin{tabular}{@{}c@{}}
      $M_{*}$ \\ ($M_\odot$)
   \end{tabular} &
   \begin{tabular}{@{}c@{}}
      $R_{*}$ \\ ($R_\odot$)
   \end{tabular} &
   \begin{tabular}{@{}c@{}}
      $T_\mathrm{eff}$ \\ (K)
   \end{tabular} &
   \multicolumn{1}{c}{$\displaystyle \frac{r_\mathrm{crit}}{a}$} &
   \multicolumn{1}{c}{$\displaystyle \frac{\Delta r}{a}^{\mathrm{a}}$} &
   \multicolumn{1}{@{}c@{}}{$\displaystyle \left(
         \frac{\tau_\mathrm{drift}}{P_\mathrm{orb}}
         \right)_\mathrm{min}^{\mathrm{b}}$} &
   \multicolumn{1}{c}{$\displaystyle \frac{\Delta r}{a}$} &
   \multicolumn{1}{@{}c@{}}{$\displaystyle \left(
         \frac{\tau_\mathrm{drift}}{P_\mathrm{orb}}
         \right)_\mathrm{min}$} &
   \multicolumn{1}{c}{$\displaystyle \frac{\Delta r}{a}$} &
   \multicolumn{1}{@{}c@{}}{$\displaystyle \left(
         \frac{\tau_\mathrm{drift}}{P_\mathrm{orb}}
         \right)_\mathrm{min}$} \\
   \hline
   \object{4U\,0115+63}4 & B0V & 16.0 & 8.0 & $3.0\,10^4$ &
        0.46 & ~0.13 & ~~2.8 & ~~0.08 & ~~1.7 & $-0.003$ & \quad -- \\
   \object{V\,0332+53} & O8.5V & 20.0 & 8.8 & $3.4\,10^4$ &
        0.50 & ~0.11 & ~~2.1 & ~~0.03 & ~~0.56 & ~~0.03 & ~~0.56 \\
   \object{A\,0535+262} & B0III & 20.0 & 15.0 & $3.1\,10^4$ &
        0.38 & ~0.09 & ~~1.2 & ~~0.05 & ~~0.66 & $-0.004$ & \quad -- \\
   \object{EXO\,2030+375} & O9III & 22.0 & 15.0 & $3.4\,10^4$ &
        0.43 &  ~0.09 & ~~1.7 & ~~0.04 & ~~0.72 & $-0.04$ & \quad -- \\
   & B0III & 20.0 & 14.0 & $2.8\,10^4$ &
        0.42 &  ~0.09 & ~~1.7 & ~~0.04 & ~~0.69 & ~~0.04 & ~~0.69 \\
   & B0V & 16.0 & 8.0 & $3.0\,10^4$ &
        0.41 &  ~0.12 & ~~2.0 & ~~0.08 & ~~1.4 & ~~0.03 & ~~0.45 \\
   \object{2S\,1417$-$624} & O9V & 20.0 & 9.0 & $3.4\,10^4$ &
        0.40 & ~0.10 & ~~1.8 & ~~0.06 & ~~1.1 & ~~0.01 & ~~0.18 \\
   & B1V & 12.0 & 7.0 & $2.5\,10^4$ &
        0.37 & ~0.11 & ~~1.9 & ~~0.08 & ~~1.4 & ~~0.04 & ~~0.75 \\
   \object{GRO\,J1008$-$57} & B0III & 20.0 & 14.0 & $2.8\,10^4$ &
        0.25 &  ~0.04 & ~~0.38 & ~~0.002 & ~~0.020 & $-0.02$ & \quad -- \\
   & B0V & 16.0 & 8.0 & $3.0\,10^4$ &
        0.24 & ~0.04 & ~~0.41 & $-0.004$ & \quad -- & $-0.03$ & \quad -- \\
   \object{2S\,1845$-$024} & B0V & 16.0 & 8.0 & $3.0\,10^4$ &
        0.084 & ~0.014 & ~~0.13 & ~~0.001 & ~~0.013 & $-0.010$ & \quad -- \\
   & B1V & 12.0 & 7.0 & $2.5\,10^4$ &
        0.081 & ~0.015 & ~~0.15 & ~~0.004 & ~~0.038 & $-0.007$ & \quad -- \\
   \hline
   \label{tab:gap}
\end{tabular}
$^{\mathrm{a}}$ $T_{\rm d} = \frac{1}{2}T_{\rm eff}$ is adopted for
all models.\\
$^{\mathrm{b}}$ $v_r = 0.1 c_\mathrm{s}$ is adopted as the upper limit
on the radial velocity. The value is given for $\Delta r > 0$.
\end{table*}

\subsection{\object{A\,0535+262}}
\label{sec:A0535+262}

This pulsar has an orbital period $P_{{\rm orb}}= 110.3\,{\rm d}$ and an
eccentricity $e =0.47$ (Finger et al. \cite{fin96a}). There is some
discussion about the exact spectral type of the optical counterpart.
Most authors support a giant in the range O9.5\,--\,B0, but Wang \&
Gies (\cite{wg98}) cannot rule out a main-sequence classification
based on UV spectra. Wang \& Gies (\cite{wg98}) constrain the mass of
the optical component to the range $8M_{\sun} \le M_{*} \le
22M_{\sun}$. If the object is a giant, the mass should be close to the
upper limit. Taking $M_{*}= 20  M_{\sun}$, the mass function indicates
$\sin i =0.43$, resulting in an orbital separation $a=286 R_{\sun}$. 

Four Type II X-ray outbursts from \object{A\,0535+262} have been
observed (in 1975, 1980, 1989 and 1994). The system also displays
series of Type I outbursts  interspersed with periods of
quiescence. The length of the series or of the quiescent states is
very variable. A series of three Type I outbursts was observed by
BATSE during 1993, followed by a Type II outburst and later two weaker
Type I outbursts (\cite{fin96a}). After that, the source has been in
quiescence and the optical counterpart has lost and reformed its
circumstellar disc (Haigh et al. \cite{nick}).

\begin{figure}[t]
   \resizebox{\hsize}{!}{\includegraphics{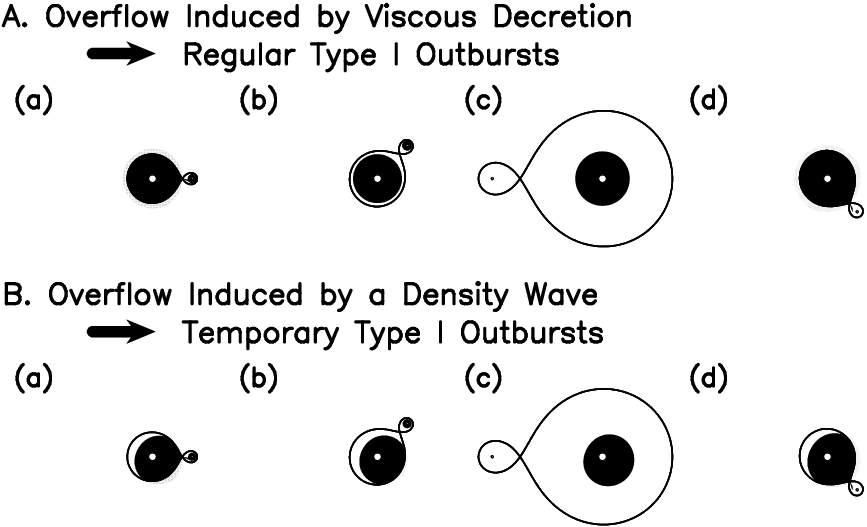}}
   \caption{ Scenarios for two families of Type~I outbursts. {\bf
       A}~Regular Type~I outbusts induced by 
     viscous decretion in an axisymmetric disc.
     (a) Near periastron, the gas continues to overflow through the $L_1$
         point, replenishing the accretion disc around the neutron
         star. [The accretion disc begins to form at panel~(d).]
     (b) At the phase at which the critical lobe radius becomes
         larger than the disc radius, the overflow stops and the
         accretion disc begins to fade.
     (c) While the neutron star orbits far from periastron,
         the disc material drifts outward by viscous decretion.
     (d) At the phase at which the critical lobe radius becomes
         smaller than the disc radius, the gas in the outermost part
         of the disc begins to overflow. The accretion disc is formed
         around the neutron star.
     {\bf B}~Temporary Type~I outbursts induced by a
     slowly precessing density wave.
     (a) If the disc is elongated roughly toward the periastron by
         chance, the overflow occurs around periastron.
     (b) At the phase at which the critical lobe radius becomes
         larger than the elongated disk size, the overflow stops
         and the accretion disc begins to fade.
     (c) Same as panel~(c) in scenario~A.
     (d) If the axis of elongation is still roughly toward the
         periastron, at the phase at which the critical lobe radius
         becomes smaller than the elongated disk size, the gas in the
         outermost part of the disc begins to overflow. The accretion
         disc is formed around the neutron star.}
   \label{fig:scenarios}
\end{figure}

Fig.~\ref{fig:alpha_c} shows that the Be disc in \object{A\,0535+262}
is expected to be truncated at the 4:1 resonance radius ($r_{\rm t}/a
\simeq 0.39$) for $0.11 \la \alpha \la 0.40$ and at the 5:1 resonance
radius ($r_{\rm t}/a \simeq 0.33$) for $0.038 \la \alpha \la 0.11$. As
shown in Fig.~\ref{fig:roche}(b) and Table~\ref{tab:gap}, the 4:1
resonance radius is slightly larger than the mean radius of the
critical lobe at periastron, while the higher resonance radii are
significantly smaller than the critical lobe. The X-ray history of
\object{A\,0535+262} suggests that the Be disc in this system has
$\alpha \ga 0.1$ and $T_{\rm d} \sim \frac{1}{2}T_{\rm eff}$ and is
truncated at the 4:1 resonance radius, which will allow enough mass
accretion on to the neutron star at every periastron passage to cause
a Type~I outburst regularly (see scenario~A in
Fig.~\ref{fig:scenarios}). If this is the case, a slight decrease in
viscosity (and/or disc temperature) will lower the disc radius 
to the 5:1 resonance radius and therefore change the system to a
dormant state, as has been repeatedly observed.

\subsection{\object{EXO\,2030+375}}
\label{sec:EXO2030+375}

This transient X-ray pulsar was discovered during a Type II outburst
in 1985. Since then, it has displayed tens of Type I X-ray outbursts
in long series separated by periods of quiescence. The orbit is
characterised by $P_{{\rm orb}}= 46.0\,{\rm d}$ and eccentricity
$e=0.41$ (Wilson et al. \cite{wil01}). The optical component of this
system is heavily obscured and no determination of the spectral type
has been possible, but both the measured mass function and the
infrared spectrum imply a spectral type earlier than about B1 (Reig et
al. \cite{rei98}). Coe et al. (\cite{coe88}) showed that the colours
of the object are compatible with a B0 spectral type. The reddening
$E(B-V)\simeq3.7$ derived by Motch \& Janet-Pacheco (\cite{mj87}) and
the distance of 5.3 kpc derived by Parmar et al. (\cite{par89}) from
the change rates in spin period and X-ray luminosity can only be
compatible if the object is a giant. However, given that the
distance determination has a large uncertainty, the possibility
  that the object is a main-sequence star cannot be ruled out.

If the spectral type is B0III, a mass of around $M_{*} = 23M_{\sun}$
is expected. We have adopted a model with $M_{*} = 20M_{\sun}$ and
$R_{*} = 14\,R_{\sun}$, in which case the mass function implies an
inclination angle $i = 56^{\circ}$ -- this is consistent with the fact
that the shape of the H$\alpha$ emission lines shown by Norton et
al. (\cite{nor94}) is typical of a Be star with moderate
inclination. The orbital separation would then be $a=150 R_{\sun}$. In
order to investigate the dependence of our model on the mass of the
primary, we have also considered the case of $M_{*} = 16\,M_{\sun}$,
$R_{*} = 8\,R_{\sun}$ (a typical B0V star).

Fig.~\ref{fig:alpha_c} shows that the difference in $\alpha_{\rm crit}$
between the B0III star model and the B0V star model is small. 
In the B0III star model, the disc is truncated at the 4:1 resonance 
radius ($r_{\rm t}/a \simeq 0.39$) for 
$0.098 \la \alpha \la 0.37$ and at the 5:1 resonance radius 
($r_{\rm t}/a \simeq 0.33$) for $0.030 \la \alpha \la 0.098$, 
while in the B0V star model the disc is truncated at the 4:1 resonance 
radius for $0.12 \la \alpha \la 0.44$ and at the 5:1 resonance radius 
for $0.036 \la \alpha \la 0.12$. We have also tried a slightly earlier
spectral type, O9III, for the primary, expecting that a higher disc
temperature would result in  a larger truncation radius. However, the
resulting radii are almost the same as those for a B0III primary,
unless $\alpha$ is as high as 0.3 (see Table~\ref{tab:gap}).

As shown in Fig.~\ref{fig:roche}(c), the 4:1 resonance radius is close
to the mean radius of the critical lobe at periastron, while the
higher resonance radii are significantly smaller than the critical
lobe. Our result suggests that the observed X-ray behaviour of
\object{EXO\,2030+375}, which has regularly exhibited Type~I X-ray
outbursts, favours a B0III or a O9III primary star with a disc with
$\alpha \ga 0.1$ and/or $T_{\rm d} \ga \frac{1}{2}T_{\rm eff}$ that is
truncated at the 4:1 resonance radius (see scenario~A in
Fig.~\ref{fig:scenarios}).

\subsection{\object{2S\,1417$-$624}}
\label{sec:2S1417-624}

\object{2S\,1417$-$624} was observed once in 1978 and was not detected
again until 1994, when a Type II outburst was observed by BATSE
(Finger et al. \cite{fin96b}). This was followed by five Type I
outbursts peaking near apastron. 

The pulsar in this Be/X-ray transient has $P_{{\rm orb}}= 42.1\,{\rm d}$
and an eccentricity $e=0.45$ (Finger et al. \cite{fin96b}). The
optical component of the system was studied by Grindlay et
al. (\cite{gri84}). Owing to the low signal-to-noise of their spectra,
they did not assign an spectral type, though several He\,{\sc ii}
lines seem to be present in the spectrum displayed, suggesting an
O-type star. Therefore we will consider again two models, one
corresponding to an O9V star ($M_{*} = 20\,M_{\sun}$, $R_{*} =
9\,M_{\sun}$) and one roughly corresponding to a B1V star ($M_{*} =
12\,M_{\sun}$, $R_{*} = 7\,M_{\sun}$).

As shown in Fig.~\ref{fig:alpha_c}, the O9V star model and the B1V
star model give rather different values of $\alpha_{\rm crit}$: the
O9V star model results in a value for $\alpha_{\rm crit}$ similar
to that in the model for \object{A\,0535+262}, whereas
the B1V star model gives a significantly higher value of $\alpha_{\rm
 crit}$. In the O9V star model, the disc is truncated at the 4:1
resonance radius ($r_{\rm t}/a \simeq 0.39$) for 
$0.14 \la \alpha \la 0.52$ and at the 5:1 resonance radius 
($r_{\rm t}/a \simeq 0.33$) for $0.047 \la \alpha \la 0.14$, 
while in the B1V star model the disc is truncated at the 4:1 resonance 
radius for $0.34 \la \alpha \la 1.2$ and at the 5:1 resonance radius 
for $0.12 \la \alpha \la 0.34$. Therefore, if the optical counterpart
of \object{2S\,1417$-$624} is a B1V star, the disc is likely to be
truncated at the 5:1 resonance radius or the 6:1 resonance radius,
which is significantly smaller than the critical lobe at periastron
(see also Table~\ref{tab:gap}). In this case, the system is unlikely
to display Type~I X-ray outbursts in a normal state, which is
consistent with the X-ray history of this system. The O9V star model
is also consistent with the X-ray history if $\alpha \la 0.1$.

We interpret the sequence of Type~I outbursts which followed the 1994
Type~II outburst as a temporary phenomenon caused by a strong
disturbance in the disc. After Type~II outbursts, the disc is expected
to be strongly disturbed (see Negueruela et al. \cite{negal01}, from
now on Paper II, for a discussion of the possible causes of Type~II
outbursts). Such a disc is likely to be strongly asymmetric and to be
temporarily capable of fuelling the neutron star (see scenario~B in
Fig.~\ref{fig:scenarios}).

\subsection{\object{GRO\,J1008$-$57}}
\label{sec:GROJ1008-57}

BATSE discovered \object{GRO\,J1008$-$57} during a Type II outburst in
July 1993. Four weak X-ray outbursts separated by $\approx 248$ days
followed. The source is believed to emit some X-ray flux during
quiescence, but no further outbursts have been observed. The optical
counterpart was identified by Coe et al. (\cite{coe94}), but no
information exists about its exact spectral type.

The orbit of this pulsar is not exactly determined. However, from the
analysis of data from the BATSE experiment, a most probable orbit can
be determined if the separation between Type I outbursts 
($247.5\,{\rm  d}$) is taken to be the orbital
period (M. Scott, priv. comm.). In that case, the eccentricity is
$e=0.66$ and $a_{{\rm x}}\sin i = 668.0 \pm 9\:{\rm lt-s}$. Again the
colours of the primary are
compatible with a B0 star. Therefore we use models for a B0V 
($M_{*} = 16\,M_{\sun}$, $R_{*} = 8\,R_{\sun}$) and a B0III ($M_{*} = 
20\,M_{\sun}$, $R_{*} = 14\,R_{\sun}$) primary.

Fig.~\ref{fig:alpha_c} shows that both stellar models give similar
values of $\alpha_{\rm crit}$. In the B0III star model, the disc is
truncated  at the 7:1 resonance radius ($r_{\rm t}/a \simeq 0.27$)
for $0.16 \la \alpha \la 0.36$ and at the 8:1 resonance radius
($r_{\rm t}/a \simeq 0.24$) for for $0.082 \la \alpha \la 0.16$. In
the B0V star model, $\alpha_{\rm crit}$ becomes higher by about 20\%.

From Fig.~\ref{fig:roche}\,(e), we immediately observe that 
the truncation radius is close to or slightly beyond the critical lobe
radius at periastron unless the viscosity is very low ($\alpha \la
0.03$). It should be noted that this is a typical feature in systems
with high orbital eccentricity, because the resonance radii are
distributed more densely for higher resonances. In other words, disc
truncation is less efficient in systems with high orbital eccentricity
than in systems with low or mild eccentricity. Moreover,
\object{GRO\,J1008$-$57} has a rather long orbital
period. Consequently, the Be disc in this system can spread out
significantly while the neutron star is far away from the Be
star. These features should enable \object{GRO\,J1008$-$57} to display
Type~I outbursts regularly.

\subsection{\object{2S\,1845$-$024}}
\label{sec:2S1845-024}

\object{2S\,1845$-$024} has been detected by several satellites at
luminosities compatible with Type I outbursts (see Finger et
al. \cite{fin99} for references). Between the launch of the {\em
ComptonGRO} satellite and 1997, BATSE detected Type I outbursts from
\object{2S\,1845$-$024} at every periastron passage. This behaviour
has continued at least until 2000 (Finger, priv. comm.).

This X-ray pulsar is in a very eccentric ($e=0.88$) and wide ($P_{{\rm
orb}} =242.2$) orbit (Finger et al. \cite{fin99}). The optical
counterpart has not been identified, but the mass function constrains
it to have $M_{*} \ga 7\,M_{\sun}$. Given that it has an exact orbital
solution, we have modelled the system in spite of the lack of data on
the optical companion. We have considered two models: one in which the
primary is our typical B0V star ($M_{*} = 16\,M_{\sun}$, $R_{*} =
8\,R_{\sun}$) -- adopting the parameters of a B0III star did not alter
significantly the value of $\alpha_{{\rm crit}}$ -- and one with
$M_{*}=12.0\,M_{\sun}$, which is close to the value used by Finger et
al. (\cite{fin99}) for their evolutionary model
($M_{*}=11.3\,M_{\sun}$), and then $R_{*}=7\,R_{\sun}$ for a typical
B1V star.

As in \object{GRO\,J1008$-$57}, both stellar models make little
difference in the value of $\alpha_{\rm crit}$
(Fig.~\ref{fig:alpha_c}), and the truncation radius is close to or
slightly beyond the critical lobe radius at periastron unless the
viscosity is very low ($\alpha \la 0.03$) [Fig.~\ref{fig:roche}(f)
and Table~\ref{tab:gap}]. With the extremely high orbital
eccentricity, the disc truncation in this system is expected to be
even less efficient than in \object{GRO\,J1008$-$57}. In addition, the
system has a long orbital period. Therefore, it is no surprise that
\object{2S\,1845$-$024} has regularly shown Type~I outbursts without
failure.

\section{Discussion}
\label{sec:discu}

In Paper~I, we analysed the disc size of \object{V635~Cas}, the
optical counterpart of \object{4U\,0115+63}, and found that the
truncated disc size is significantly smaller than the distance to the
first Lagrangian ($L_1$) point at periastron. Among our conclusions,
we found that it is precisely this wide gap between the disc outer
radius and the position of the $L_1$ point that prevents the system
from displaying Type~I outbursts, by drastically reducing the
accretion rate on to the neutron star. In Paper~II we pointed out
that, for the very same reason, the density in the disc of
\object{V635~Cas} is likely to grow with time, to a point where the
disc becomes optically thick and unstable to the radiation-driven
warping, which seems to be at the origin of Type~II X-ray outbursts.

As discussed in the end of Sect.~\ref{sec:trunc}, we cannot expect a
perfect truncation. Since the disc density is expected to decrease
rapidly beyond the truncation radius, the gap size determines how
effective the disc truncation is. For systems with a wide gap, like
\object{4U\,0115+63}, the truncation is so effective that the system
does not show Type~I outbursts under normal conditions. Whatever
little amount of disc material manages to overcome the tidal
truncation may be easily ejected from the vicinity of the neutron star
by the propeller effect (Stella et al. \cite{swr86}). On the other
hand, for systems with a narrow gap, the truncation will not be
efficient, allowing the neutron star to capture gas from the disc at
every periastron passage and display Type~I outbursts.

Therefore, the main purpose of this paper is to find out the general
trend in the dependence of the gap size on various parameters. A
second purpose is to understand why we can see two different kinds of
Type~I outburst series: regular series of outbursts of similar
intensity in some systems and short series of outbursts of varying
intensity (generally associated with Type II outbursts) in other
systems.

\subsection{Systems with large eccentricity}

We have found that, in systems with high orbital eccentricity, say $e
\ga 0.6$, such as \object{GRO\,J1008$-$57} ($e=0.66$) and
\object{2S\,1417$-$624} ($e=0.88$), the truncation radius of the Be
disc is very close to, in the sense that
$(\tau_\mathrm{drift})_\mathrm{min} \ll P_\mathrm{orb}$,
or slightly beyond the critical lobe radius at
periastron unless the viscosity is very low ($\alpha \la 0.03$). Under
such conditions, we expect disc truncation not to be efficient,
allowing the accretion by the neutron star of enough mass at every
periastron passage to cause a Type~I X-ray outburst regularly (see
scenario~A in Fig.~\ref{fig:scenarios}). Moreover, the long orbital
period of \object{GRO\,J1008$-$57} ($P_{\rm orb}=247.5\,{\rm d}$) and
\object{2S\,1845$-$024} ($P_{\rm orb}=242.2\,{\rm d}$) will enable the
Be disc in these systems to spread out significantly while the neutron
star is far from periastron. This larger drift time will allow the
neutron star in such systems to accrete rather more matter at
periastron passage than in a system with shorter orbital period.

Among systems which we expect to be relatively well described by these
conditions, we list \object{4U\,1258$-$61} (\object{GX 304$-$1}) which
displayed regular Type I outbursts every $132.5\:{\rm d}$ during most
of the 1970s (Corbet et al. \cite{coral86}) and which has been dormant
ever since because of the disappearance of the Be disc. Similarly, the
behaviour of \object{4U\,1145$-$619}, which displays short and not
very strong outbursts every $\approx188\:{\rm d}$, is indicative of a
large eccentric orbit. This is supported by the lower limit $e\la0.6$
found by Cook \& Warwick (\cite{cw87}) from the analysis of changes in
the X-ray pulse timing. Occasionally, some of the periodic outbursts
from \object{4U\,1145$-$619} are rather stronger (challenging the
conventional definition of Type II outbursts, since they cannot be
described as very bright). This may be related to a large global
perturbation in the Be disc, reflected in the profile of emission
lines (Stevens et al. \cite{stev97}). \object{RX J0812.4$-$3114} has
also shown a long series of Type I outbursts without any intervening
Type II outburst during a 4-year quasi-cycle of Be disc formation and
dissipation (Reig et al. \cite{rei01}). Finally
\object{SAX\,J2239.3+6116}, which has shown several Type I outbursts
separated by $\approx 262\:{\rm d}$, could also belong to this
category.

One very peculiar system is \object{A\,0535$-$668}, which has a very
short orbital period ($P_{\rm orb}=16.7\,{\rm d}$) and is believed to
have a very high eccentricity (Charles et al. \cite{char83}). If the
eccentricity is really $e\ga0.8$, the neutron star passes close to the
surface of the Be star at periastron. Our model is simply not
applicable to such an extreme system.

It must be noted that the observed behaviour of
\object{GRO\,J1008$-$57} is only partially compatible with the orbital
model. The Type~I outbursts are short and not very strong, with
similar luminosities, as expected for such a high-eccentricity
system. However, the fact that only four have been observed and that
they occurred after a Type~II outburst casts some doubts about the
validity of the orbital solution used.

\subsection{Systems with low eccentricity}

It is well known
that accretion discs in circular binaries with mass ratio $0.05 \la q
\la 0.25$ are truncated at the 3:1 resonance radius by the
tidally driven eccentric instability (e.g., Osaki \cite{osa96}). Since the
mass ratios of all Be/X-ray binaries fall within this range and the
mechanism of truncation is the same for decretion discs as for
accretion discs, we expect that
Be discs in systems with very low orbital eccentricity, say $e \la
0.2$, are also truncated at the 3:1 resonance radius.
Consequently, the gap size in these systems should be much wider than in 
systems with larger eccentricities. Wide gaps in
low-eccentricity systems will result in disc truncation being so effective
that no Type~I outbursts should occur unless the Be disc is very
strongly disturbed. Therefore it is likely that systems with very low
orbital eccentricity will show only Type~II outbursts (and perhaps
temporarily Type~I outbursts only when the disc is strongly disturbed).

From this point of view, \object{GS\,0834$-$430} is an extraordinary system,
since it has rather low eccentricity (constrained to be
$0.10\la e\la0.17$), but has shown a long series of
Type~I outbursts (Wilson et al. \cite{wil97}). It must be noted that the
abrupt change in the separation between outbursts observed during this
series is highly
unusual. Moreover, the decrease in peak intensity of the outbursts
along the series suggests that the Be disc was strongly
perturbed at the time of the outbursts. We have to admit, however,
that with the information available, our model obviously fails to
explain this system. 

Apart from \object{X Per}, which shows no outbursts, systems with low
eccentricity are \object{XTE\,J1543$-$568} ($e< 0.03$), which has been detected
only recently and does not seem to show any clear modulation of its
X-ray lightcurve at the orbital period (in't Zand et al. \cite{zan01}), and
2S\,1553$-$542 ($e< 0.09$), which was only detected once in 1975
(Kelley et al. \cite{kel83}). Our
interpretation of their behaviour is that these objects, because of
effective disc truncation, never display Type I outbursts. Their discs
therefore accumulate mass until very large perturbations develop. Only
occasionally will one of these perturbations result in transfer of
matter towards the neutron star, producing a bright Type II outburst.

In this respect, it is necessary to insist that the physical
mechanisms leading to Type II outbursts are not well understood. Our
model indicates that the truncated discs in Be/X-ray binaries
cannot reach a steady state and will continuously grow denser. As a
consequence, we expect the discs to become dynamically unstable. In
the case of \object{4U\,0115+63}, which has been carefully studied
(and perhaps also in \object{A\,0535+262}), there is observational
evidence that links big dynamical instabilities and Type II outbursts
(Paper II). From this observational fact, we conclude that the
dynamical perturbation {\em somehow} results in the transfer of large
amounts of material from the Be disc to the neutron star. No aspect of
our model implies that such mass transfer {\em must} occur. It may
well be that in systems with low eccentricity, dynamical instability
generally leads to the collapse of the disc and its fallback on to the
Be star without any substantial amount of material reaching the
neutron star. If this is the case, low-eccentricity transients would
be completely undetectable most of the time.

Until recently, it had been generally assumed that the orbits of
Be/X-ray binaries were in general rather eccentric owing to the effect
of supernova kicks. The recent determination of orbital parameters for
several systems with low eccentricity (Delgado-Mart\'{\i} et
al. \cite{del01}; in't Zand et al. \cite{zan01}) may represent
a challenge to this idea. If there is a selection effect against the
detection of systems with low eccentricities, their population could
be rather larger than previously suspected. If eccentric systems
display X-ray outbursts rather more frequently than systems with low
eccentricities, they would be more likely to be detected
over a limited time-span (such as the time since the start of X-ray
astronomy).

The orbit of the pulsar \object{GRO\,J1948+32} is constrained to have
an eccentricity $e\la0.25$ and an orbital period $35\: {\rm d}<P_{{\rm
 orb}}<70\: {\rm d}$ (Chakrabarty et al. \cite{chak95}). Its recent
identification with the Be/X-ray transient \object{KS\,1947+300} means
that it has probably displayed at least three Type II outbursts
between 1989 and 2001. An accurate determination of its orbit would
therefore be very important in understanding whether  this selection
effect really exists. If its eccentricity turns out to be very low, 
then we will have to conclude that low-eccentricity systems may
display Type II outbursts as frequently as systems with moderate
eccentricities. If, in contrast, its eccentricity is close to
$e=0.25$, the available data would suggest that systems with low
eccentricities have less frequent Type II outbursts. 

A second system that could serve to understand the outbursting
behaviour of systems with low eccentricities is the LMC transient
{\object{EXO\,0531.1$-$6609}}. Based on a statistical parameter
study, Dennerl et al. (\cite{den95}) conclude that this pulsar
($P_{{\rm s}}=13.7\:{\rm s}$) is likely to have an orbital period 
$P_{{\rm orb}}=25.4\:{\rm d}$ and low eccentricity
$e\approx0.1$. {\object{EXO\,0531.1$-$6609}} has displayed at least
three relatively bright outbursts extending over more than one orbital
period (i.e., Type II) in 1983, 1985 and 1993, but also seems to have
been occasionally detected by {\em ROSAT} at a low X-ray luminosity
close to periastron. Obviously confirmation of the orbital parameters
is necessary before any conclusions can be drawn.

\subsection{Systems with moderate eccentricity}

Four systems (\object{V\,0332+53}, \object{A\,0535+262},
\object{EXO\,2030+375}, \object{2S\,1417$-$624})
among the six Be/X-ray binaries discussed in the previous section have
moderate eccentricity. As we have already seen, the model results
for these systems turn out to depend on rather subtle details, in
contrast to the much more robust predictions for systems with high
eccentricity and very low eccentricity. Since the spacing of the
candidate truncation radii is rather larger in these systems than in
high-eccentricity Be/X-ray binaries, a small difference in the system
parameters  
(orbital, stellar or disc parameters) can produce significantly
different X-ray behaviour, depending on whether the resulting
truncation radius happens to be close to the critical lobe radius or
not. Therefore we have not tried to present general predictions for
these systems. Instead, in the previous section we have estimated the
truncation radii and system parameters by comparing the
model results with the observed X-ray behaviour. In this sense, the 
current status of our model can still be considered to be ``at the
qualitative level'' for  mildly eccentric systems. We estimated that
Be discs in \object{V\,0332+53}, \object{A\,0535+262} in the X-ray
active state, and \object{EXO\,2030+375} are truncated at the 4:1
resonance radius, while those in \object{A\,0535+262} in quiescence
and \object{2S\,1417$-$624} are truncated at the 5:1 resonance
radius. The smaller truncation radius for \object{A\,0535+262} in
quiescence might suggest a slightly lower disc temperature in this
state than in the X-ray active state. To understand more about these
systems, we need a more sophisticated model for the interaction
between the Be disc and the neutron star.

It is worth noting, however, that all the intermediate
eccentricity Be/X-ray binaries known have displayed Type II
outbursts. This is the case for the four systems listed here and also
for \object{GRO\,J1750$-$27}, a presumably very distant source (with
$P_{\rm {orb}} = 29.8\:{\rm d}$ and $e=0.36$) which has only been
detected once during a Type II outburst (Scott et
al. \cite{sco97}). This fact suggests that disc truncation is
generally rather effective for orbital eccentricities $e\la 0.5$, even
though the discs in \object{A\,0535+262} and \object{EXO\,2030+375},
under certain conditions, seem to reach a state in which truncation is
not so effective and long series of Type I outbursts occur (not
surprisingly, these two systems have the largest eccentricities and
longest periods among the five known).

Though systems with moderate eccentricities dominate the sample of
Be/X-ray binaries for which orbital solutions exist at present, it is
important to note that at least two important selection effects may
have been at work in the definition of this sample:

\begin{enumerate}
\item Be/X-ray binaries with moderate
   eccentricities are much more likely to display Type II outbursts than
   systems with high eccentricities and perhaps also more likely than
   systems with low eccentricities (see above). Their detection as
   new X-ray sources is therefore easier.
   \newline\hspace*{1em}
   Among Be/X-ray binaries, \object{4U\,0115+63} has been by far the most
   active during the era of X-ray astronomy, displaying no fewer than
   13 Type II outbursts between 1969 and 2000. From the estimates in
   Paper II, we come to the conclusion that the enhanced activity of
   \object{4U\,0115+63} is not owing to causes intrinsic to the Be
   star (such as a particularly high mass-loss rate) and therefore it
   is likely to be connected to its orbital parameters. 
   \newline\hspace*{1em}
   The combination of moderate eccentricities and relatively close
   orbits results in rather efficient disc truncation, which allows
   the storage of matter in the Be disc, but at the same time results
   in gaps which are not too wide and may permit the transfer of part
   of this material to the neutron star under certain conditions.

\item Once the system has been detected, a relatively close orbit
  may be solved from X-ray observations spanning one single Type II
  outbursts, while in a larger orbit a much longer baseline will be
  needed (only recently, permanent coverage over long time-spans with
  BATSE and {\em RXTE} has allowed the solution of some orbits with
  $P_{{\rm orb}} \ga 50\:{\rm d}$).
\end{enumerate}

Because of these two factors, judgement on what sort of X-ray
behaviour is more ``typical'' among Be/X-ray binaries has to be
suspended until a rather larger sample is known.

\subsection{Predictions}

Many of the systems under discussion have shown short series of
irregular Type~I outbursts. This category includes \object{4U\,0115+63},
\object{V\,0332+53}, \object{2S\,1417$-$624} and presumably 
\object{GRO\,J2058+42}, for which no
orbital solution exists. These series have generally occurred in
connection with Type~II outbursts (in most cases, after a Type~II
outburst). The behaviour of \object{A\,0535+262} has also been
similar on some occasions. As mentioned in
Sect.~\ref{sec:2S1417-624}, we expect that the
temporary Type~I outbursts occur in systems in which the Be disc has
a radius significantly smaller than (but not too small compared to)
the critical lobe radius under normal conditions. These systems will
be mostly mildly eccentric systems. If the Be disc in such a
system is strongly disturbed by, e.g., radiation-driven warping and/or 
a global density wave, it will become strongly asymmetric. Then, if the
disc elongation is roughly in the periastron direction, the disc
will be capable of fuelling the neutron star and cause a sequence of
Type~I X-ray outbursts (see scenario~B in Fig.~\ref{fig:scenarios}).
Since \object{GRO\,J2058+42} has apparently displayed
similar behaviour, we expect it to have moderate eccentricity.

A relationship between the X-ray behaviour and the orbital
eccentricity stems naturally from the truncated disc model. In
general, the presence of regular Type~I X-ray outbursts indicates that
the truncated disc around the Be star has a radius close to or
slightly larger than the critical lobe radius at periastron. Such a
situation occurs in all systems with highly eccentric orbit, some
mildly eccentric systems, and no systems with low eccentricity
(\object{GS\,0834$-$430} may be an exception). In a rough sense, the
lower the orbital eccentricity, the wider the gap between the
truncation radius and the critical lobe radius and therefore the more
difficult the accretion on to the neutron star. 

A source which hardly fits in this picture is
\object{A\,0726$-$26}. This persistent low-luminosity X-ray source
which has never shown outbursts displays, however, a 35-d modulation
in its X-ray emission (Corbet \& Peele \cite{cp97}). If this is the
orbital period, \object{A\,0726$-$26} presents very few similarities
to other Be/X-ray binaries, falls very far away from the
$P_{{\rm s}}/P_{{\rm orb}}$ correlation and cannot be explained by our
model.

Among the Be/X-ray binaries with known counterparts, two systems have
been very rarely detected as X-ray sources, and even then they have
been relatively weak. \object{A\,1118$-$616} has only been detected
twice, in 1975 and in 1991. In both cases, the outburst was long, but
not very bright. \object{Cep X-4} has been detected four times between
1972 and 1997. All the outbursts have also been relatively weak. The
two sources display permanent low-luminosity X-ray emission when not
in outburst. In both cases, the counterparts present evidence for very
large circumstellar discs. Therefore we expect these two systems to
have wide and not very eccentric orbits.

We note that even at the
qualitative level our model has some predictive power and can thus be
tested. Finding out, for example, that \object{RX J0812.4$-$3114} has
an orbital eccentricity close to zero or that  \object{A\,1118$-$616}
has a very eccentric orbit would certainly force us to reconsider most
of the conclusions reached here.

\section{Concluding Remarks}

We have applied the resonantly truncated disc model developed by
Negueruela \& Okazaki (\cite{no01}) to six systems which have
displayed Type~I X-ray outbursts.
 We have found that the model 
naturally explains the X-ray behaviour of these systems and 
that all systems are consistent with having a similar viscosity parameter
$\alpha \sim 0.1$.

In our model, regular Type~I X-ray outbursts occur in all
systems with high orbital eccentricity, some mildly eccentric systems,
and no systems with low eccentricity. On the other hand, systems which
have temporarily shown Type~I outbursts will be mostly mildly
eccentric systems, in which the Be disc temporarily becomes strongly
asymmetric.

In the scenario presented here for Be/X-ray transients, disc material
can reach the neutron star only via the $L_1$ point, and will
therefore have a not very high velocity relative to the neutron
star. Since such a flow (which may be said to represent the disc
version of Roche-lobe overflow) carries high angular momentum, an
accretion disc may temporarily be formed around the neutron star
during each X-ray outburst. Such temporary accretion discs have
probably been observed in \object{2S\,1845$-$024} (Finger et
al. \cite{fin99}), but extensive searches have failed to detect them
around \object{A\,0535+262} (Motch et al. \cite{mot91}). The accretion
flow in the vicinity of the neutron star (i.e., once it has left the
effective Roche lobe of the Be star) should be modelled in detail in
order to understand if such discs should always form and to predict
the X-ray lightcurves derived from our scenario.

\begin{acknowledgements}
We thank John~M.~Porter, Mark~H.~Finger, and Gordon Ogilvie for many
helpful discussions.
We also thank the referee Marten H. van Kerkwijk for his valuable
comments.
ATO thanks the Institute of Astronomy, Cambridge,
UK, and the Observatoire de Strasbourg, France, for the warm
hospitality.
\end{acknowledgements}


\end{document}